\documentclass[sigconf,acmtog,screen,nonacm]{acmart}
\authorsaddresses{}
\author{Zehao Xia}
\email{5445687d@gmail.com}
\affiliation{%
  \institution{Chongqing University}
  \country{China}
}

\author{Yiqun Wang}
\authornote{Corresponding author.}
\email{csyqwang@hotmail.com}
\affiliation{%
  \institution{Chongqing University}
  \country{China}
}

\author{Zhengda Lu}
\email{luzhengda@ucas.ac.cn}
\affiliation{%
  \institution{University of Chinese Academy of Sciences}
  \country{China}
}

\author{Kai Liu}
\email{liukai0807@gmail.com}
\affiliation{%
  \institution{Chongqing University}
  \country{China}
}

\author{Jun Xiao}
\email{xiaojun@ucas.ac.cn}
\affiliation{%
  \institution{University of Chinese Academy of Sciences}
  \country{China}
}

\author{Peter Wonka}
\email{pwonka@gmail.com}
\affiliation{%
  \institution{KAUST}
  \country{Saudi Arabia}
}

\AtBeginDocument{%
  }
  
\acmJournal{TOG}
\acmSubmissionID{2004}
\usepackage{enumitem}
\usepackage[table]{xcolor}
\usepackage{booktabs}
\usepackage{amsmath}
\usepackage{multirow} 
\usepackage{pifont}

\usepackage{xcolor}

\definecolor{gold}{RGB}{255, 178, 178}
\definecolor{silver}{RGB}{255, 217, 178}

\begin{document}

\newcommand{\method}{$\Omega$-Avatar}
\title{OMEGA-Avatar: One-shot Modeling of $360^\circ$ Gaussian Avatars}

\begin{abstract}
Creating high-fidelity, animatable 3D avatars from a single image remains a formidable challenge. We identified three desirable attributes of avatar generation: 1) the method should create avatars from a single image {in a one-shot setting without per-identity network optimization}, 2) model a $360^\circ$ full-head, and 3) produce animation-ready results. However, current works address only two of the three points simultaneously.
To address these limitations, we propose OMEGA-Avatar, the first framework for one-shot reconstruction of animatable, 360°-complete full-head avatars.
Starting from a one-shot animatable avatar framework, we achieve $360^\circ$ full-head avatar generation with two novel components.
First, to overcome poor hair modeling in full-head avatar generation, we introduce a semantic-aware mesh deformation module integrating multi-view normals to optimize a FLAME head with hair while preserving its topology.
Second, to enable effective feed-forward decoding of full-head features, we propose a multi-view feature splatting module constructing a shared canonical UV representation from features across multiple views through differentiable bilinear splatting, hierarchical UV mapping, and visibility-aware fusion. 
This approach preserves both global structural coherence and local high-frequency details across all viewpoints, ensuring $360^\circ$ consistency without per-instance {network} optimization. 
Extensive experiments demonstrate that OMEGA-Avatar achieves state-of-the-art performance, significantly outperforming baselines in $360^\circ$ full-head completeness while {maintaining competitive cross-view identity consistency.}
\end{abstract}



\begin{teaserfigure}
  \centering
  \includegraphics[width=\textwidth]{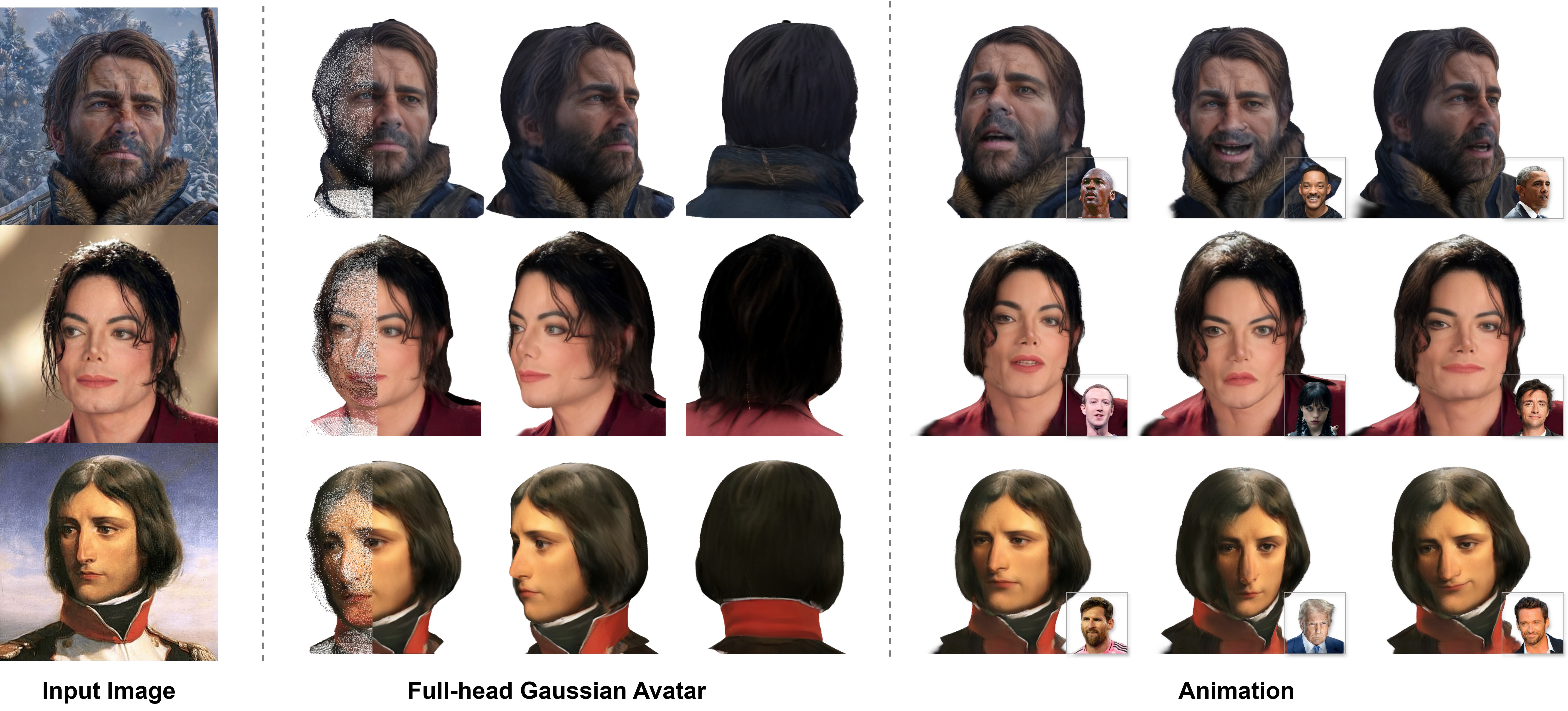}
  \caption{Given a single portrait image across diverse styles (left), our method {performs one-shot reconstruction of a high-fidelity full-head Gaussian avatar} (middle). The reconstructed avatars include the back of the head and are capable of high-quality animations (right).}
  \Description{teaser}
  \label{fig:teaser}
\end{teaserfigure}

\maketitle

\section{Introduction}
\begin{figure*}[t]
  \centering
  \includegraphics[width=\textwidth]{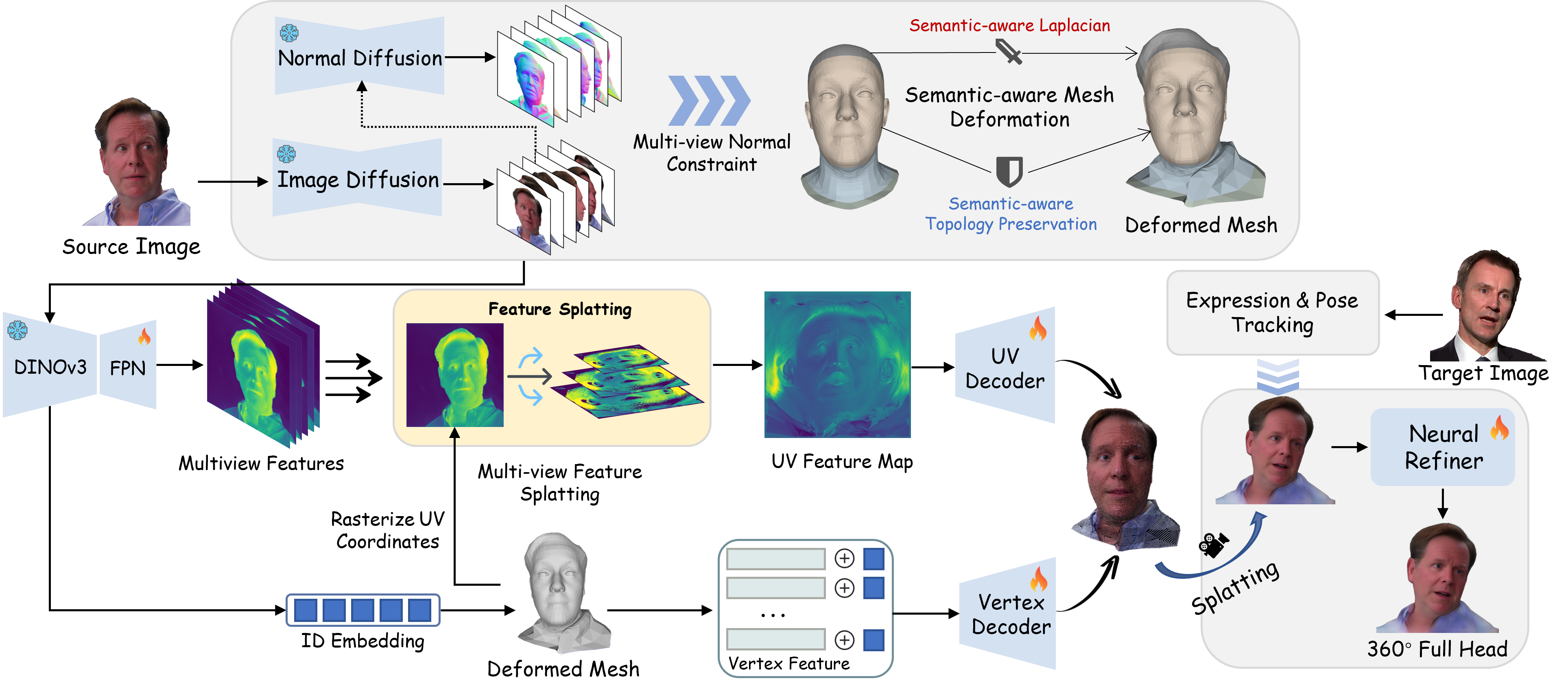}
  \caption{\textbf{Pipeline Overview.} Given the source and target images, we leverage diffusion models to synthesize multi-view RGB images and corresponding normal maps. These normal maps are used for semantic-aware mesh deformation, while pixel-wise features are extracted from multi-view RGB images. Multi-view features are subsequently aggregated into a canonical UV feature map through the multi-view feature splatting module. The UV features and vertex features extracted from the deformed mesh are decoded and anchored to the mesh via UV mapping. For animation, the expression and pose derived from the target image are injected into the deformed mesh. Finally, the rendered output is enhanced by a neural refiner to generate the final full-head avatar.}
  \Description{A schematic pipeline diagram illustrating the full-head avatar generation process. The diagram shows source and target images as inputs, followed by diffusion-based synthesis of multi-view RGB images and normal maps. The normal maps guide mesh deformation, while features extracted from the multi-view images are aggregated into a canonical UV feature map through a feature splatting module. The UV features and mesh vertex features are decoded and mapped onto the deformed mesh. Expression and pose information from the target image are then applied to the mesh for animation, and the final rendered result is refined by a neural refinement module.}

  \label{fig:pipeline}
\end{figure*}

Generating high-fidelity, animatable 3D full-head avatars from a single image is a pivotal challenge in computer graphics. To make this under-constrained problem tractable, we identify three essential properties the avatar generator must possess. 
First, it should generalize to unseen identities without per-identity network optimization at test time. 
Due to the under-constrained single-image input, recovering complete and animatable full-head avatars inevitably relies on strong learned priors from large-scale data. 
This motivates a one-shot formulation that generalizes across identities without optimizing model parameters for each subject, unlike approaches requiring costly per-subject network tuning. 
Second, it must support $360^\circ$ full-head modeling. Unlike standard facial reconstruction, complete avatars require inferring occluded geometry and appearance, like hair and the back of the head, ensuring geometric completeness. This typically demands multi-view supervision to hallucinate reasonable content in unseen regions. 
Third, the avatar should be animation-ready. Beyond static geometry, models must support explicit control over expressions and poses. This relies on modeling additional motion-related information through parametric models or video-based supervision, ensuring reconstructed avatars are consistently driven by explicit motion parameters while preserving identity.
However, satisfying all three attributes remains elusive. Generative approaches~\cite{panohead, spherehead, RODIN} leverage 3D-aware GANs or diffusion models for $360^\circ$ head synthesis. However, lacking explicit parametric head representations, they remain constrained by inversion-based or sampling-heavy workflows, hindering efficient inference and animatable head generation. While FaceLift~\cite{FaceLift} attempts frame-by-frame animation via diffusion, its stochastic nature leads to temporal inconsistency and jittering. Methods like FATE~\cite{FATE} and SOAP~\cite{soap} achieve animatable full heads by leveraging multi-view priors, but rely on instance-specific pipelines requiring costly optimization or remeshing, limiting their efficiency and generalizability. Recent one-shot methods~\cite{GAGAvatar, LAM} train feed-forward networks on large-scale 2D video datasets~\cite{VFHQ}. While excelling at frontal reenactment, they cannot decode full-head features in a feed-forward manner, causing structural distortions in side and back views and failing to generate $360^\circ$ full-head avatars.
To bridge these gaps, we propose \method{}, the first framework for one-shot reconstruction of animatable, 360°-complete full-head avatars. 
Following GAGAvatar~\cite{GAGAvatar}, we use one-shot to denote reconstructing an animatable 3D Gaussian head avatar from a single image without any per-identity network optimization at test time.
Within this {one-shot} and animation-ready formulation, we propose two novel components that tackle full-head avatar generation (Fig.~\ref{fig:pipeline}). First, to address the difficulty of modeling complete head geometry with limited 3D datasets, we leverage a pre-trained multi-view diffusion model to augment large-scale video data, producing consistent multi-view normal maps and RGB images without multi-view capture. Synthesized normal maps serve as geometric priors to optimize a FLAME mesh with hair, enforcing cross-view consistency while preserving underlying topology. Second, a key challenge is effectively fusing discrete and scale-inconsistent multi-view features into a unified representation for feed-forward decoding. We thus propose a multi-view feature splatting mechanism mapping features from generated multiple views into a shared canonical UV space. This UV-based representation serves as a differentiable bridge, enabling robust aggregation of global structure and local high-frequency details across viewpoints. The resulting UV features are decoded into 3D Gaussians and bound to the FLAME mesh via UV mapping. Jointly leveraging these components enables \method{} to generate high-fidelity animatable avatars {without per-identity network optimization at test time}. Our main contributions are as follows.

\begin{itemize}[leftmargin=*, noitemsep, topsep=0pt, parsep=0pt, partopsep=0pt]
\item We propose the first one-shot framework generating high-fidelity, full-head, and animatable avatars from a single image.
\item We introduce a multi-view feature splatting module, a fully differentiable multi-view feature fusion mechanism, constructing a shared canonical UV feature map from discrete 2D multi-view features for 3D Gaussian decoding.
\item Extensive experiments demonstrate that our method achieves state-of-the-art performance on multi-view full-head generation, outperforming the main baseline (GAGAvatar) in both full-head completeness and multi-view visual fidelity on both datasets.
\end{itemize}

\begin{table}[t]
\centering
\caption{\textbf{Comparison of properties with state-of-the-art methods.}
Our method is the first framework for one-shot reconstruction of animatable
$360^\circ$ full-head avatars with 3D Gaussian Splatting.
}
\label{tab:feature_comparison}

\resizebox{\linewidth}{!}{
\begin{tabular}{lccccc}
\toprule
Method & Representation & One-Shot  &
$360^\circ$ Full-Head & Animatable \\
\midrule

Rodin \cite{RODIN} 
& Tri-plane 
& \ding{51} 
& \ding{51} 
& \ding{55} \\

PanoHead \cite{panohead} 
& Tri-plane 
& \ding{55}$^*$ 
& \ding{51} 
& \ding{55} \\

SphereHead \cite{spherehead} 
& Tri-plane 
& \ding{55}$^*$ 
& \ding{51} 
& \ding{55} \\

FaceLift \cite{FaceAdapter} 
& 3DGS 
& \ding{51} 
& \ding{51} 
& \ding{55} \\

\midrule

FATE \cite{FATE} 
& 3DGS 
& \ding{55} 
& \ding{51} 
& \ding{51} \\

GAGAvatar \cite{GAGAvatar} 
& 3DGS 
& \ding{51} 
& \ding{55} 
& \ding{51} \\

LAM \cite{LAM} 
& 3DGS 
& \ding{51} 
& \ding{55} 
& \ding{51} \\

SOAP \cite{soap} 
& Mesh 
& {\ding{51} (Slow Recon. Speed)}
& \ding{51} 
& \ding{51} \\

\midrule

\textbf{Ours} 
& \textbf{3DGS} 
& \textbf{\ding{51}} 
& \textbf{\ding{51}} 
& \textbf{\ding{51}} \\

\bottomrule
\end{tabular}
}

\vspace{1mm}

\parbox{\linewidth}{
\raggedright
\footnotesize
$^*$ Denotes methods requiring per-instance GAN inversion prior to reconstruction.
}
\end{table}

\section{Related work}
\noindent \textbf{Single-Image Head Avatar Reconstruction.}
Early 3DMM-based approaches \cite{ROME, Deep3DPortrait} provide strong statistical priors but lacked fine-grained details. With the advent of Neural Radiance Fields~\cite{NeRF} (NeRF), the field shifts toward implicit representations, offering higher fidelity in modeling complex attributes like hair~\cite{Nerfies, HFA-GP, PointAvatar, INSTA, Export3D}. Most NeRF-based methods require identity-specific multi-view or video data, limiting generalization to unseen subjects and raising privacy concerns. To bypass large-scale videos, generative approaches~\cite{pi-gan, StyleNeRF, StyleSDF} and 3D-aware GANs like EG3D~\cite{EG3D} achieve high visual quality via efficient tri-plane representations. However, reconstructing specific identities requires costly latent inversion~\cite{PTI, HFGI3D}, often degrading accuracy and failing to fully preserve identity. Addressing these limitations, recent works target one-shot 3D head reconstruction~\cite{HideNeRF, OTAvatar, NOFA, Real3D-Portrait, GOHA, CVTHead, LCNF, GPAvatar, portrait4d, portrait4dv2, voodoo3d, LSF-Animation}. 
3D Gaussian Splatting (3DGS)~\cite{3DGS} offers real-time photorealistic rendering and drives recent one-shot adaptations~\cite{GAGAvatar, LAM}; {these methods often deteriorate under large viewpoint changes because they primarily focus on frontal portrait animation rather than complete $360^\circ$ full-head reconstruction.
Beyond head avatars, recent Gaussian reconstruction models predict 3D Gaussians from single or sparse-view images~\cite{SplatterImage, PixelSplat, MVSplat, GRM, GS-LRM}. LGM~\cite{LGM}, the closest analogue to our pipeline, combines multi-view diffusion with Gaussian regression for generic 3D content generation.  Unlike these generic reconstruction methods, our method constructs FLAME-bound Gaussians in a canonical head space, enabling explicit expression- and pose-driven animation.
Related Gaussian head models include GGHead~\cite{GGHead}, an unconditional 3D Gaussian head generator rather than a single-image identity reconstruction method, and Avat3r~\cite{avat3r}, which primarily reconstructs animatable heads from multi-view posed input images. Their main problem settings are therefore not directly aligned with our single-image $360^\circ$ full-head reconstruction setting.}
Our work addresses these shortcomings by proposing a robust framework that maintains high rendering efficiency and geometric consistency for one-shot 3D full-head animation. 

\noindent \textbf{Generative Full-Head Reconstruction.}
Recent full-head modeling methods rely on generative models for novel view synthesis. GAN-based approaches~\cite{panohead,spherehead} enable high-quality $360^\circ$ head synthesis via volumetric representations. Diffusion-based approaches~\cite{RODIN,RodinHD} generate triplane head representations but remain static and unsuitable for animation.
SOAP~\cite{soap} reconstructs textured, mesh-based full-head avatars from single portraits, but relies on iterative FLAME deformation and adaptive remeshing, resulting in costly per-input reconstruction and requiring rigging attributes to be transferred after topology changes.
To address these limitations, we present a novel framework capable of efficient one-shot full-head reconstruction supporting high-quality animation.

\section{Methodology}
\subsection{Overview}
We present a framework for generating high-fidelity, animatable full-head Gaussian avatars from a single image, as illustrated in Fig.~\ref{fig:pipeline}.
We leverage diffusion models to generate RGB images and normal maps, guiding semantic-aware FLAME mesh deformation to obtain a personalized mesh.
Our dual-branch architecture generates Gaussians that construct the 3D representation. The UV Gaussian branch encodes appearance details, employing DINOv3~\cite{DINOv3} with a Feature Pyramid Network (FPN) to extract pixel-aligned features that fuse deep semantics with high-frequency geometric details. These multi-view features are aggregated into a unified UV feature map via the splatted UV feature map module. The vertex Gaussian branch anchors Gaussians directly to the FLAME vertices to ensure structural coherence.
Finally, Gaussians from both branches are driven by the underlying FLAME model {and deformed according to the target expression and pose parameters. A 2D neural refiner is used during training and for the refined animation setting. For novel-view synthesis, we use raw 3D renderings to preserve cross-view consistency.}
In the following subsections,
{Sec.~\ref{sec:stage1}} describes the process of semantic-aware mesh deformation. 
{Sec.~\ref{sec:stage2}} elaborates on the canonical full-head avatar construction, introducing the specifics of both the UV and vertex Gaussian branches. 
\subsection{Semantic-aware Mesh Deformation.}
\label{sec:stage1}
We leverage the pretrained diffusion model from SOAP~\cite{soap}, built upon the Unique3D~\cite{unique3d} framework, to synthesize six RGB images $\mathbf{I}_{rgb}$ and normal maps $\mathbf{I}_{nml}$ from the input.
Given the generated multi-view consistent normal maps $\mathbf{I}_{nml}$ and facial landmarks detected using~\cite{landmark}, we aim to reconstruct a personalized mesh while maintaining a clean topology suitable for animation.
Unlike previous FLAME tracking methods~\cite{vhap} relying solely on landmarks or single-view photometric constraints, our approach is supervised by multi-view consistent normal maps. This rich geometric prior allows us to recover intricate 3D details typically lost in parametric fitting.
To recover the personalized geometry, we propose a single-stage optimization framework. In contrast to prior art~\cite{soap} that requires multi-stage iterative remeshing to handle deformations, our method efficiently deforms the mesh in {a single optimization stage without remeshing.}
We initialize the mesh using the FLAME template and introduce a deformation field. Let $\mathcal{M}(\beta, \theta, \psi)$ denote the standard FLAME driven by shape $\beta$, pose $\theta$, and expression $\psi$. We optimize a per-vertex offset $\Delta \mathbf{V} \in \mathbb{R}^{N \times 3}$ to capture high-frequency geometric details, where $N$ denotes the number of vertices. Deformed vertices $\mathbf{V}'$ are formulated as:
\begin{equation}
\mathbf{V}' = \mathcal{M}(\beta, \theta, \psi) + \Delta \mathbf{V}.
\end{equation}
\begin{figure}[t]
  \centering
  \includegraphics[width=0.8\columnwidth]{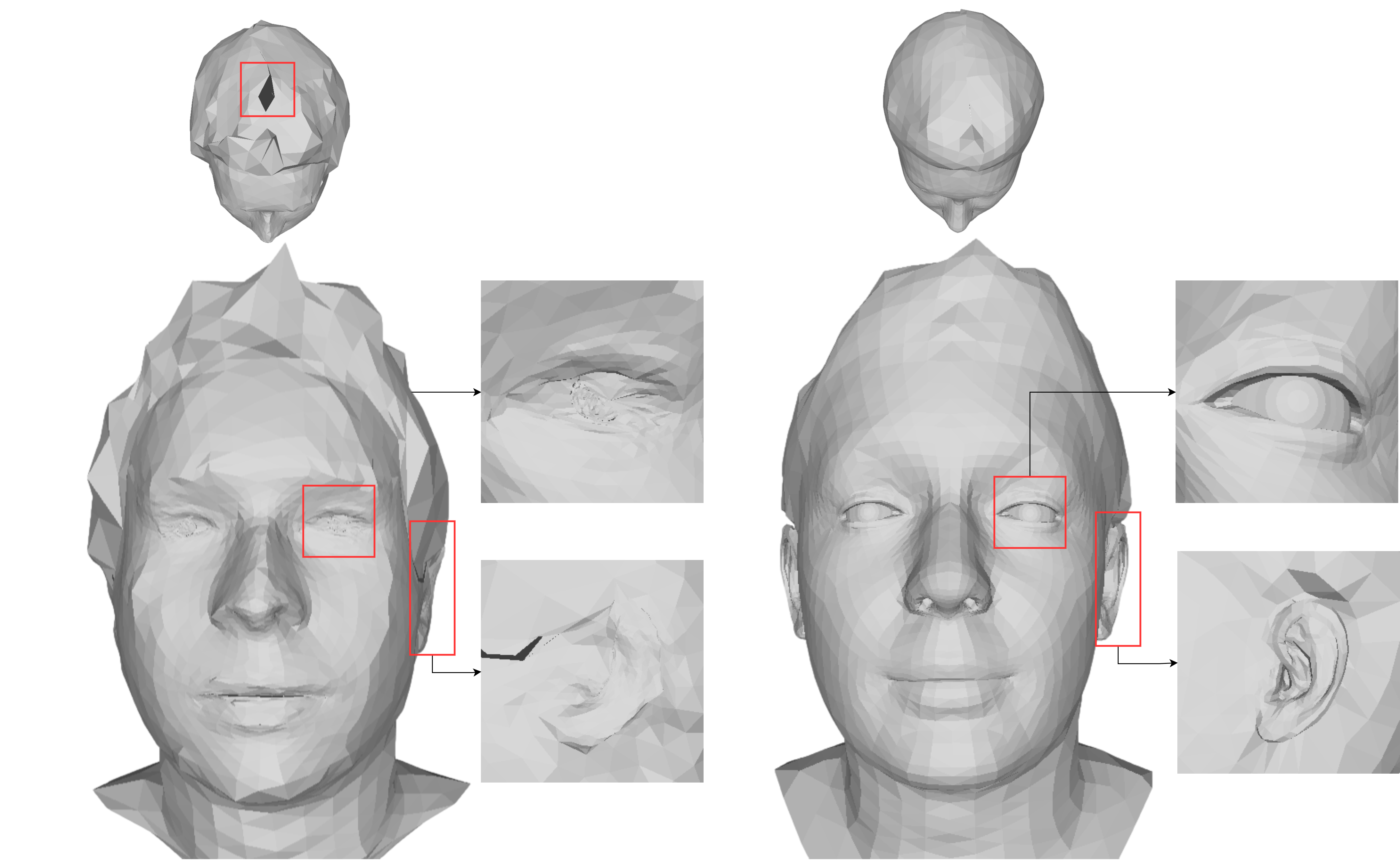}
  \caption{\textbf{Semantic-aware Mesh Deformation.} Direct optimization with normal guidance disrupts FLAME's parametric structure, causing severe surface irregularities and topological artifacts (left). Note the cranial holes and degenerated facial features like eyes and ears (highlighted in red boxes). Our approach (right) mitigates this via semantic-aware topology preservation and a semantic-aware Laplacian. We preserve FLAME's clean topology and ensure $360^\circ$ geometric consistency, enabling fine detail generation without compromising facial structural integrity.}
  \Description{}
  \label{fig:lap}
\end{figure}
\noindent \textit{Semantic-aware Topology Preservation.}
A core challenge in mesh deformation is balancing geometric fidelity with animatability. Unconstrained optimization of $\Delta \mathbf{V}$ across the entire mesh often disrupts the inherent semantic topology of the FLAME model, degrading expression reenactment quality. To address this, we introduce a semantic-aware optimization strategy. Specifically, we utilize a semantic mask to restrict the normal-guided deformation. We apply the normal consistency constraints strictly to the non-facial regions while preserving the parametric structure of the facial region. This design disentangles the optimization of static geometry from dynamic facial components, ensuring high-fidelity animation.

To ensure accurate and topologically coherent deformation, we optimize $\Delta \mathbf{V}$ by minimizing the following composite objective function:
\begin{equation}
\mathcal{L} = \lambda_{nml}\mathcal{L}_{nml} + \lambda_{lmk}\mathcal{L}_{lmk} + \lambda_{lap}\mathcal{L}_{lap},
\end{equation}
The primary supervision comes from the normal consistency loss $\mathcal{L}_{nml}$, which minimizes the $L_2$ distance between the rendered normal of the deformed mesh and the diffusion-predicted maps $\mathbf{I}_{nml}$. To ensure semantic alignment, we employ a landmark loss $\mathcal{L}_{lmk}$ that penalizes the projection error of 3D facial landmarks and enforces symmetry in the canonical space. 

\noindent \textit{Semantic-aware Laplacian.} Although restricting normal constraints to non-facial regions helps preserve facial topology, the cranial region remains susceptible to surface irregularities and topological artifacts lacking top-view supervision, as illustrated in Fig.~\ref{fig:lap}. Standard Laplacian regularization is insufficient here: applying a uniform weight strong enough to smooth the cranial artifacts would inadvertently over-smooth the facial features. To resolve this conflict, we introduce a semantic-aware Laplacian smoothing term $\mathcal{L}_{lap}$ with a spatially-varying weighting scheme:
\begin{equation}
\mathcal{L}_{lap} = \sum_{i=1}^{N} {w}_i \| \delta_i \|^2, \quad \delta_i = {v}'_i - \frac{1}{|{N}_{nei}(i)|} \sum_{j \in {N}_{nei}(i)} {v}'_j,
\end{equation}
where $\delta_i$ is the Laplacian coordinate approximation for vertex $i$. Consistent with our semantic-aware strategy, the weight $w_i$ is not constant; we assign significantly higher weights to the hair and boundary regions to suppress degeneration, while assigning minimal weights to the facial region. This ensures the cranium remains smooth without compromising the high-frequency details of the face preserved by the parametric prior.
\begin{figure}[t]
  \centering
  \includegraphics[width=\columnwidth]{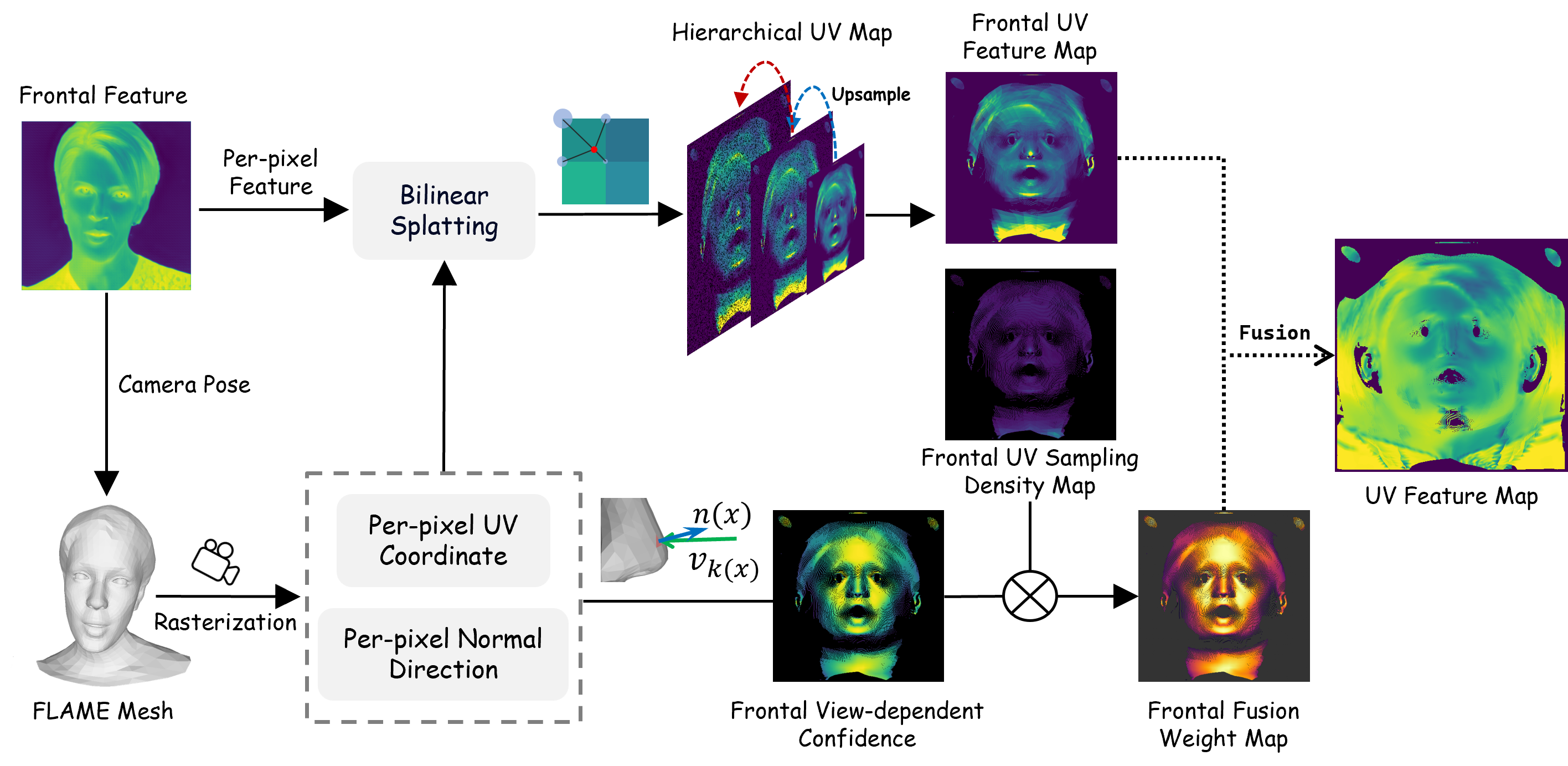}
  \caption{\textbf{Multi-view Feature Splatting.} Taking a frontal view as an example, we first obtain per-pixel UV coordinates and normals via rasterization, mapping features to UV space using differentiable bilinear splatting. Our hierarchical UV mapping builds a multi-resolution pyramid to fill missing regions in a coarse-to-fine manner. Simultaneously, we calculate a fusion weight map by combining view-dependent confidence and UV sampling density. Finally, the visibility-aware fusion module aggregates weighted features across all views to generate the final UV feature map.}
\Description{Diagram illustrating the per-view feature processing pipeline. For each input view, rasterization yields per-pixel UV coordinates, followed by bilinear splatting of 2D features onto hierarchical UV maps. The resulting partial UV feature maps are merged via visibility-aware fusion to form a complete canonical UV feature map.}
  \label{fig:MFS}
\end{figure}

\subsection{Canonical Gaussian Full-head Avatar Reconstruction}
\label{sec:stage2}  
We model the canonical full-head avatar with a hybrid Gaussian representation that combines vertex Gaussians for expression control and UV Gaussians for view-consistent appearance modeling.

The Vertex Gaussian Branch anchors 3D Gaussians to FLAME vertices to enable explicit expression driving. We extract a global identity token $f_{id}$ from the frontal view, as it provides the most comprehensive identity information. Simultaneously, to distinguish different vertices and encode spatial details, we assign a unique learnable parameter to each vertex $v^{(i)}$. The global token is concatenated with these local parameters and fed into a vertex decoder to predict the Gaussian attributes $\mathcal{G}_{\text{vert}} = \{ r^{(i)}, s^{(i)}, \alpha^{(i)}, c^{(i)} \}$. This design ensures geometric stability during animation.

\begin{figure*}[t] 
  \centering
  \includegraphics[width=\textwidth]{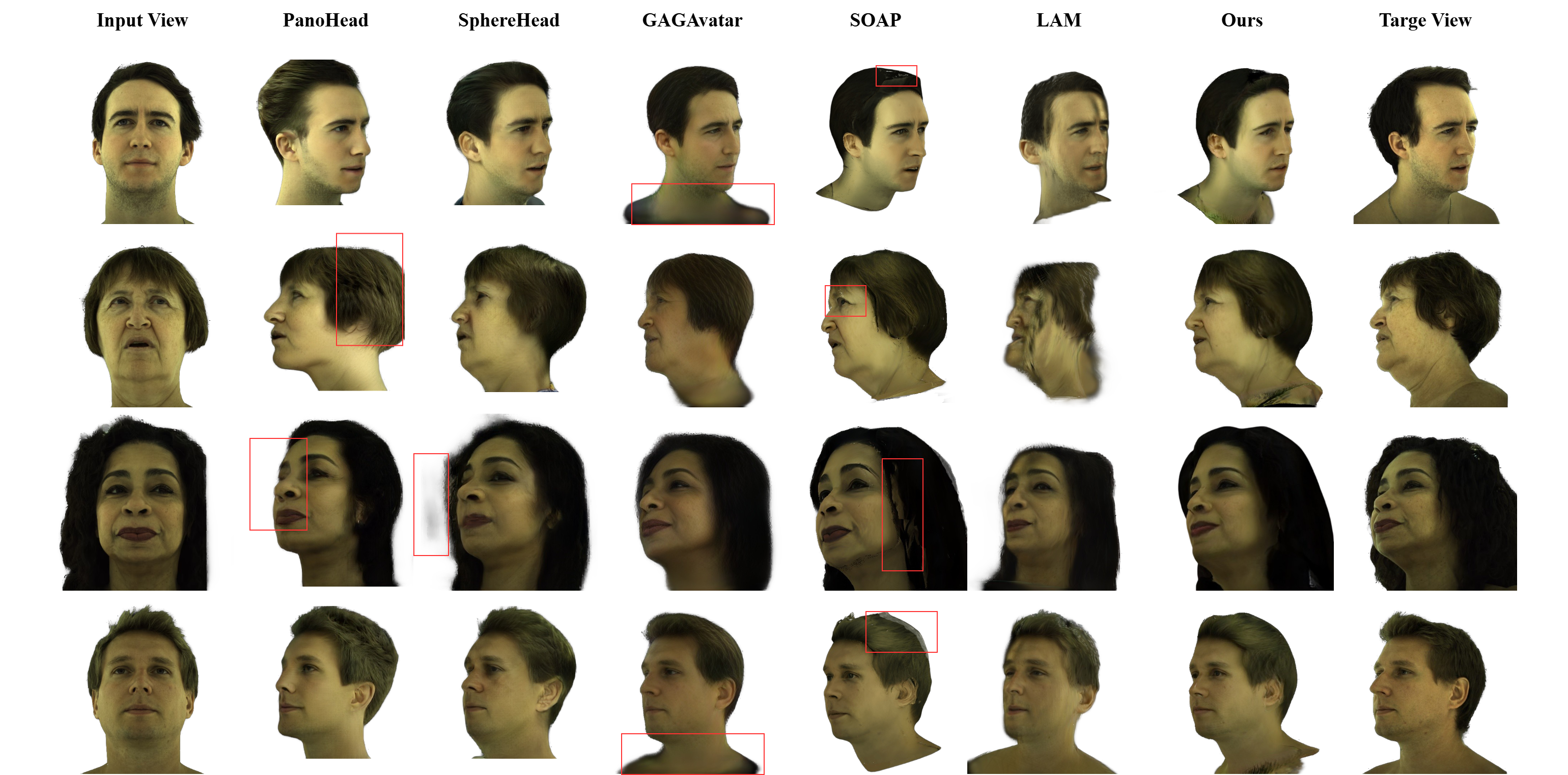}
  \caption{\textbf{Novel view synthesis from single image on the Avatar-256 dataset.} Compared to state-of-the-art methods, our approach produces high-quality rendering results, even under unseen and extreme side-view facial angles. Note that PanoHead and SphereHead require inputs aligned to the FFHQ canonical space, leading to apparent scale differences. We use the red boxes to highlight the visual artifacts.}
\Description{A qualitative comparison showing rendered head avatars from multiple methods under extreme side-view facial angles. The visual results illustrate differences in identity consistency and rendering quality across methods, with the proposed approach producing more stable facial structure and fewer visual artifacts in challenging viewpoints.}
\label{fig:qualitative}
\end{figure*}

While the vertex branch ensures structural stability, the limited number and discrete nature of FLAME vertices restrict vertex Gaussians from representing high-frequency details. To capture appearance while maintaining surface topological continuity, we introduce the UV Gaussian Branch. Given $k$ generated multi-view images, we employ an image encoder to extract semantic feature maps $\mathbf{F}_k$.
However, effectively aggregating multi-view feature maps is non-trivial due to view-dependent variations and inconsistent feature scales. To achieve this, we propose a \textbf{Multi-view Feature Splatting} module that lifts multi-view 2D features onto a canonical UV map, enabling joint decoding of all views in a single pass. Our approach consists of three integral components: differentiable bilinear splatting, hierarchical UV mapping, and visibility-aware fusion, as shown in Fig.~\ref{fig:MFS}. This enables stable end-to-end gradient propagation from multi-view feature maps to the UV feature map.

\noindent \textbf{Differentiable Bilinear Splatting.}
We first establish dense correspondences between screen pixels and UV coordinates.
For a given view $k$ with camera pose $\mathbf{P}_k$, we perform rasterization to retrieve a continuous UV coordinate $\mathbf{u}_p \in \mathbb{R}^2$ for each pixel $p$ in the feature map.
Direct nearest-neighbor assignment is prone to quantization artifacts. Instead, we adopt differentiable bilinear splatting and formulate it as a differentiable aggregation operator over the UV domain.
Given a pixel feature $\mathbf{f}_p$ and its continuous UV coordinate $\mathbf{u}_p$, bilinear splatting softly distributes the feature to neighboring UV grid locations according to a bilinear kernel induced by $\mathbf{u}_p$.

Aggregating contributions from all pixels yields a per-view UV feature map $\mathbf{U}_k$ and a corresponding density map $\mathbf{D}_k$:
\begin{equation}
\mathbf{U}_k = \sum_{p} \mathcal{B}(\mathbf{u}_p)\, \mathbf{f}_p,
\qquad
\mathbf{D}_k = \sum_{p} \mathcal{B}(\mathbf{u}_p),
\end{equation}
where $\mathcal{B}(\mathbf{u}_p)$ denotes the bilinear splatting operator induced by $\mathbf{u}_p$.
$\mathbf{D}_k$ records the effective sampling density in the UV space.

\newcommand{\first}[1]{%
    \setlength{\fboxsep}{1.5pt}\colorbox{gold}{{#1}}}
\newcommand{\second}[1]{%
    \setlength{\fboxsep}{1.5pt}\colorbox{silver}{#1}}
\newcommand{\third}[1]{%
    \setlength{\fboxsep}{1.5pt}\colorbox{bronze}{#1}} 

\begin{table*}[t] 
    \centering
    \caption{\textbf{Quantitative Comparison of Novel View Synthesis on the Avatar-256 and NeRSemble Datasets.} Colors denote the \first{best} and \second{second-best}}
    \label{tab:quantitative}
    \resizebox{\textwidth}{!}{
    \begin{tabular}{lccccc ccccc}
        \toprule
        & \multicolumn{5}{c}{\textbf{Avatar-256 Novel Views}} & \multicolumn{5}{c}{\textbf{NeRSemble Novel Views}} \\ 
        \cmidrule(lr){2-6} \cmidrule(lr){7-11} 
        Method & PSNR $\uparrow$ & SSIM $\uparrow$ & LPIPS $\downarrow$ & CSIM $\uparrow$ & DS $\downarrow$ & PSNR $\uparrow$ & SSIM $\uparrow$ & LPIPS $\downarrow$ & CSIM $\uparrow$ & DS $\downarrow$ \\
        \midrule
      
        PanoHead    & 17.995 & 0.5932 & 0.2592 & 0.4556 & 0.1663 & 17.151 & 0.6098 & 0.2706 & 0.5742 & 0.1082 \\
      
        SphereHead  & 18.932 & 0.6238 & 0.2436 & 0.4714 & \first{0.1509} & 17.446 & 0.6183 & 0.2628 & 0.5781 & 0.1053 \\
      
        GAGAvatar   & \second{22.802} & \second{0.7714} & \second{0.1719} & 0.4682 & 0.1762 & \second{22.556} & \second{0.8027} & \second{0.1505} & \first{0.6867} & \second{0.1033} \\
      
        SOAP        & 21.335 & 0.7287 & 0.1948 & \first{0.5783} & 0.1890 & 19.539 & 0.7471 & 0.1975 & \second{0.6855} & 0.1128 \\
      
        LAM         & 21.802 & 0.7314 & 0.1997 & 0.4842 & 0.1818 & 20.263 & 0.7465 & 0.1853 & 0.6321 & 0.1118 \\
      
        Ours        & \first{23.244} & \first{0.7734} & \first{0.1592} & \second{0.5403} & \second{0.1651} & \first{23.221} & \first{0.8051} & \first{0.1435} & 0.6714 & \first{0.0950} \\
        \bottomrule
    \end{tabular}
    }
\end{table*}

\noindent \textbf{Hierarchical UV Mapping.}
A critical challenge in splatting is the resolution mismatch between screen space and texture space, which often results in sparse coverage and "holes" in the UV map.
To address this issue, we introduce a hierarchical UV map representation. Instead of splatting features only at the target UV resolution, we splat them onto a pyramid of UV maps with progressively lower resolutions.
At each pyramid level $l$, bilinear splatting produces a feature map $\mathbf{U}_{k,l}$ and an associated density map $\mathbf{D}_{k,l}$.
To synthesize the final high-resolution map, we employ a coarse-to-fine hole-filling mechanism. Coarser UV maps, which possess broader receptive fields and denser coverage, are upsampled to the target resolution. We fuse these upsampled features with the finer-scale features using a soft, differentiable mask derived from the density map $\mathbf{D}_{k,l}$. Specifically, regions with low splatting density at the fine level are smoothly filled by the upsampled global context from coarser levels, effectively repairing artifacts without introducing discontinuities.

\noindent \textbf{Visibility-aware Fusion.}
After obtaining the completed UV feature maps from one view, we fuse the multi-view UV feature maps into a single canonical UV feature map. The fusion weight for each view is defined by combining three complementary factors.

View Weight: Since frontal views typically contain richer information, we assign each view a predefined global weight $\gamma_k$.

View-dependent Confidence: We define a view-dependent geometric confidence to measure the reliability of observations under a given view $k$. For each pixel $p$, the included angle between the surface normal $\mathbf{n}_p$ and the viewing direction $\mathbf{v}_k$ determines the reliability of the feature. We aggregate these per-pixel confidence scores into the UV domain using the same splatting operator:
\begin{equation}
\mathbf{C}_k = \sum_{p} \mathcal{B}(\mathbf{u}_p) \cdot  \max\!\left(0,\, \mathbf{n}_p \cdot \mathbf{v}_k\right).
\end{equation}
This naturally down-weights grazing-angle observations while preserving strong signals from front-facing surfaces.

UV Sampling Density: The accumulated density $\mathbf{D}_{k,l}$ reflects how many features contribute to a given UV coordinate at pyramid level $l$, serving as a measure of feature reliability.
The final canonical UV feature map $\mathbf{U}$ is obtained by a globally normalized weighted average over all views $k$ and pyramid levels $l$. We define the composite fusion weight map $\mathbf{W}_{k,l}$ as the element-wise product of view weight $\gamma_k$, the geometric confidence $\mathbf{C}_k$, and the sampling density $\mathbf{D}_{k,l}$:
\begin{equation}
\mathbf{W}_{k,l} =\gamma_k \cdot \mathbf{C}_k \odot \mathbf{D}_{k,l}.
\end{equation}
Using these weights, the fused feature map is computed as:
\begin{equation}
\mathbf{U} = \frac{\sum_{k} \sum_{l} \mathbf{W}_{k,l} \odot \mathbf{U}_{k,l}}{\sum_{k} \sum_{l} \mathbf{W}_{k,l} + \epsilon}.
\end{equation}
This formulation prioritizes densely observed and front-facing regions, while allowing coarser-scale features and alternative views to smoothly fill missing or weakly observed areas.

We train the framework using a self-reenactment scheme, minimizing the discrepancy between the rendered output and the driving frame. The total objective function is a composite of reconstruction, perceptual, multi-view constraints, and local regularization (details in Supplementary Material).

\section{Experiments}
\subsection{Experimental Setting}
We compare with five recent works: PanoHead \cite{panohead}, SphereHead~\cite{spherehead}, GAGAvatar~\cite{GAGAvatar}, LAM~\cite{LAM}, and SOAP~\cite{soap}. PanoHead and SphereHead are state-of-the-art methods for 3D full-head reconstruction using 3D GANs. GAGAvatar and LAM are one-shot animatable avatar reconstruction methods based on 3D Gaussian Splatting. SOAP is a diffusion-based method for reconstructing the full head by remeshing a FLAME mesh. For evaluation metrics, please refer to the Supplementary Material. 
{Regarding computational cost, on a single NVIDIA RTX 4090, mesh fitting takes approximately 30~s per input, feed-forward Gaussian decoding takes 468.24~ms once per avatar. FLAME expression driving and rendering take 7.73~ms and 13.60~ms per frame, respectively, resulting in a total animation cost of 21.33~ms per frame (46.88~FPS), with 6.6 GB of GPU memory.}

\begin{figure*}[t] 
  \centering
  \includegraphics[width=\textwidth]{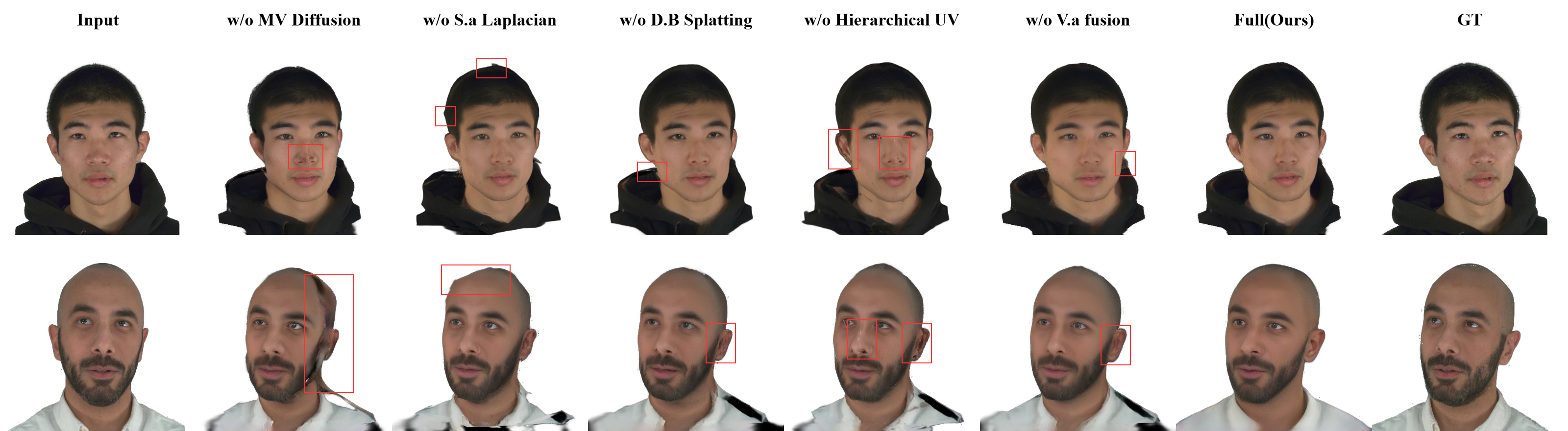}
  \caption{\textbf{Qualitative Ablation on the NeRSemble dataset.} We compare results for models trained: (1) without multi-view diffusion, (2) without using the semantic-aware Laplacian term, (3) without differentiable bilinear splatting, (4) without hierarchical UV mapping, (5) without visibility-aware fusion.}
  \Description{Ablation}
  \label{fig:ablation}
\end{figure*}

\subsection{Main Results}
To evaluate our method's 3D consistency and novel view synthesis, we leverage two public multi-view datasets: NeRSemble~\cite{nersemble} and Avatar-256~\cite{avatar-256}. For both datasets, reconstruction is performed from a single input image, while multi-view observations are used only for evaluation; full details of the dataset splits, view selection, and camera settings are provided in the supplementary material. All novel-view synthesis results are evaluated using raw 3D renderings without the 2D neural refiner.
For animation quality, we conduct expression transfer experiments on VFHQ~\cite{VFHQ} and evaluate on the held-out HDTF~\cite{HDTF} dataset to assess cross-dataset generalization, comparing against SOAP~\cite{soap}, the leading $360^\circ$ animatable baseline. 
{Additional comparisons with the frontal-oriented one-shot animatable baselines GAGAvatar\cite{GAGAvatar} and LAM\cite{LAM} are provided in the supplementary material, while SOAP is used as the primary baseline here due to its more closely matched $360^\circ$ full-head setting.}
To further assess generalization, we collect in-the-wild face images for qualitative evaluation. 

\noindent\textbf{Qualitative results.}
Fig.~\ref{fig:qualitative} presents qualitative comparisons against baselines under novel viewpoints. Unlike competing approaches, our method synthesizes photorealistic 3D renderings while maintaining consistency in fine-grained details and identity, even under extreme viewpoint variations. 
In contrast, PanoHead and SphereHead struggle to preserve identity across views, suffering from severe identity-view ambiguity. Constrained by single-view training data and 3D priors, GAGAvatar and LAM frequently exhibit structural distortions at large viewing angles. Regarding the mesh-based SOAP, its remeshing process is prone to surface fragmentation, leading to visible artifacts. Overall, our method improves multi-view visual fidelity while retaining competitive identity consistency.

\begin{figure*}[t] 
  \centering
  \includegraphics[width=0.97\textwidth, height=0.97\textheight, keepaspectratio]{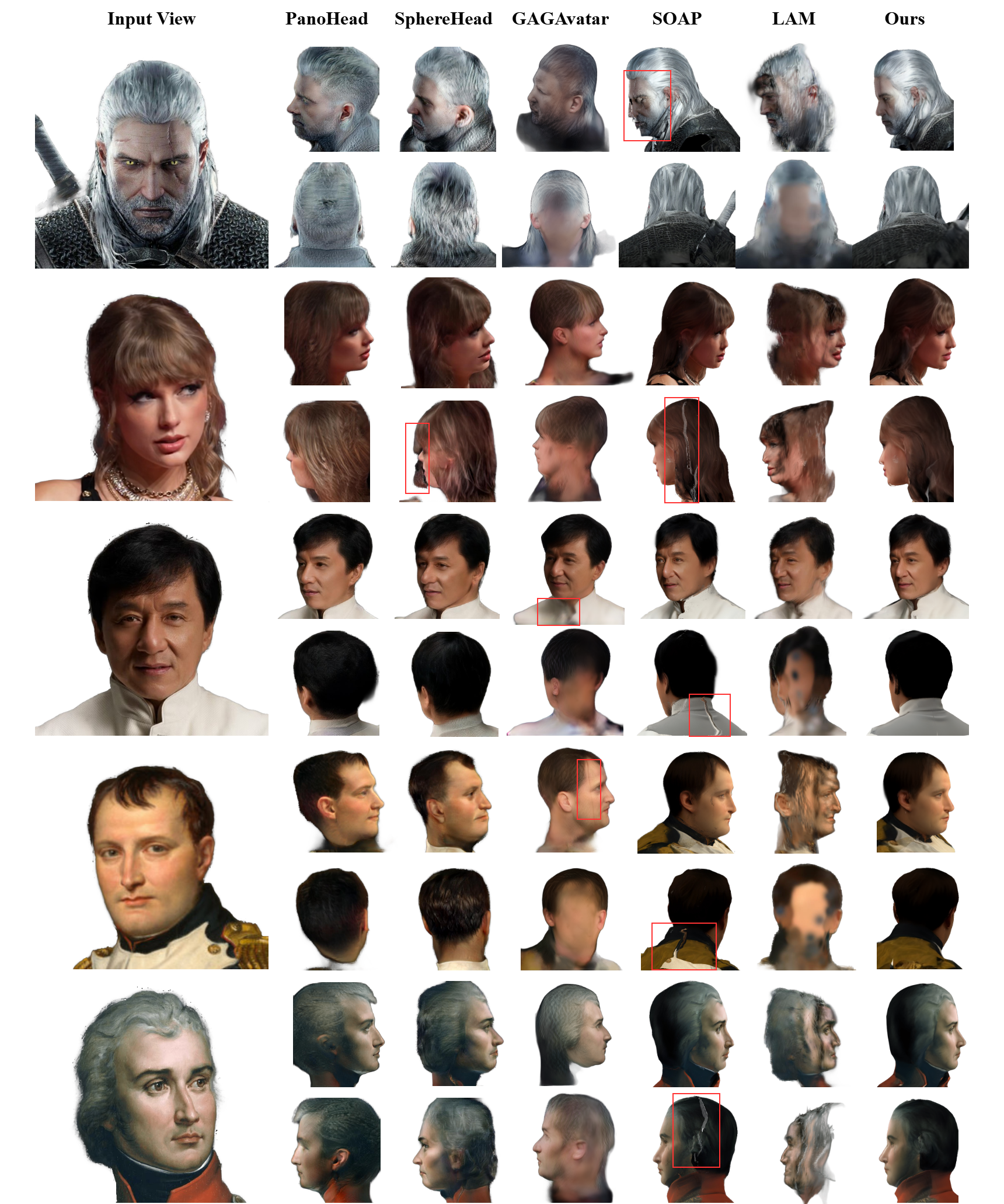}
  \caption{\textbf{Additional Results on In-the-wild Images.} Our method demonstrates great generalization ability and robustness towards in-the-wild images, and provides realistic unseen view rendering results. Red boxes highlight visual artifacts.}
  \Description{}
  \label{fig:In-the-wild}
\end{figure*}

\noindent \textbf{Quantitative results.}
Tab.~\ref{tab:quantitative} reports quantitative results on the NeRSemble and Avatar-256 datasets, respectively. For visual fidelity, our method demonstrates superior rendering quality as evidenced by the PSNR, SSIM, LPIPS, and DreamSim (DS) metrics, {while achieving competitive CSIM. CSIM is averaged across all novel viewpoints; however, as the underlying ArcFace features are less reliable under large-yaw and back views, we interpret CSIM together with the other perceptual metrics rather than as a standalone measure of cross-view identity consistency.} Regarding reconstruction efficiency, PanoHead and SphereHead necessitate time-consuming Pivotal Tuning Inversion to optimize latent codes; SOAP relies on a multi-stage remeshing process for each subject; {GAGAvatar and LAM use feed-forward Gaussian prediction, while our Gaussian decoder operates feed-forward after the preceding multi-view synthesis and mesh-fitting stages.}
The former two baselines lack multi-view priors, resulting in suboptimal performance on multi-view datasets compared to our approach. Ultimately, our method realizes one-shot, animatable full-head reconstruction, achieving state-of-the-art performance on both the NeRSemble and Avatar-256 benchmarks.
{Tab.~\ref{tab:reenactment} and Tab.~\ref{tab:reenactment_hdtf} compares SOAP with our raw and refined animation outputs. Here, Ours (w/o refiner) denotes the raw output of the 3D Gaussian renderer without applying the 2D neural refiner, whereas Ours (full) denotes the corresponding refined output. Our method consistently outperforms SOAP across the vast majority of metrics, demonstrating superior expression fidelity during animation. Additional qualitative stress tests under high-intensity facial expressions are provided in the supplementary material.

\begin{table}[htbp]
    \centering
    \footnotesize
    \setlength{\tabcolsep}{4.7pt}
    \caption{\textbf{Reenactment evaluation on the VFHQ dataset.}}
    \label{tab:reenactment}
    \resizebox{\columnwidth}{!}{%
    \begin{tabular}{lccccccc}
        \toprule
        Method & PSNR $\uparrow$ & SSIM $\uparrow$ & LPIPS $\downarrow$ & CSIM $\uparrow$ & AED $\downarrow$ & APD $\downarrow$ & AKD $\downarrow$ \\
        \midrule
        SOAP & 14.557 & 0.599 & 0.305 & 0.662 & 0.175 & 0.0562 & 4.997 \\
        Ours (w/o refiner) & 18.062 & 0.714 & 0.188 & \textbf{0.664} & 0.116 & 0.0207 & 2.196 \\
        Ours (full) & \textbf{20.739} & \textbf{0.744} & \textbf{0.169} & 0.660 & \textbf{0.109} & \textbf{0.0188} & \textbf{2.023} \\
        \bottomrule
    \end{tabular}%
    }
\end{table}


\begin{table}[htbp]
    \centering
    \footnotesize
    \setlength{\tabcolsep}{4.7pt}
    \caption{\textbf{{Reenactment evaluation on the HDTF dataset.}}}
    \label{tab:reenactment_hdtf}
    \resizebox{\columnwidth}{!}{%
    \begin{tabular}{lccccccc}
        \toprule
        Method & PSNR $\uparrow$ & SSIM $\uparrow$ & LPIPS $\downarrow$ & CSIM $\uparrow$ & AED $\downarrow$ & APD $\downarrow$ & AKD $\downarrow$ \\
        \midrule
        SOAP & 14.607 & 0.711 & 0.283 & 0.735 & 0.166 & 0.0351 & 3.579 \\
        Ours (w/o refiner) & 19.854 & 0.765 & 0.166 & \textbf{0.757} & 0.107 & \textbf{0.0090} & 1.551 \\
        Ours (full) & \textbf{22.594} & \textbf{0.820} & \textbf{0.123} & 0.752 & \textbf{0.097} & 0.0091 & \textbf{1.515} \\
        \bottomrule
    \end{tabular}%
    }
\end{table}
}
\noindent \textbf{Results on In-the-wild Images.}
To further evaluate generalization, we collected a set of in-the-wild facial images. Fig.~\ref{fig:In-the-wild} presents qualitative comparisons against baselines in these unconstrained scenarios. PanoHead and SphereHead often fail to accurately reconstruct complex hairstyles. GAGAvatar and LAM suffer from severe structural distortions in back and side views. While SOAP recovers accurate global geometry, it often exhibits visible cracks in the back of the head due to mesh fragmentation. These results collectively demonstrate the robust generalization and superior full-head reconstruction capabilities of our method.

\subsection{Ablation Studies}
We conduct ablation studies on single-view reconstruction using the NeRSemble dataset. Quantitative and qualitative results are presented in Tab.~\ref{tab:ablation} and Fig.~\ref{fig:ablation}.

\noindent \textbf{w/o MV Diffusion.}
We remove the multi-view diffusion step. Significant artifacts appear at the rendering boundaries of the frontal view. Furthermore, the synthesis quality for side and rear views degrades drastically, failing to reconstruct a complete, animatable 3D avatar. This confirms the necessity of diffusion-based priors for hallucinating unobserved geometry and texture.

\noindent \textbf{w/o Semantic-aware Laplacian.}
We remove the semantic-aware Laplacian constraint. As shown in Fig.~\ref{fig:ablation}, this leads to geometric inconsistencies. The rendering fidelity decreases, with noticeable surface irregularities emerging along the head and facial contours, resulting in bumpy geometry, indicating that the Laplacian term is crucial for maintaining surface smoothness and mesh integrity.

\noindent \textbf{w/o Differentiable Bilinear Splatting.}
We replace our differentiable bilinear splatting with a hard nearest-neighbor assignment and fix the density weights to 1. Although the LPIPS metric shows a slight improvement—likely due to absent interpolation blur—all other quantitative metrics decline. Qualitatively, the lack of differentiable gradients and soft distribution leads to quantization artifacts and fine-grained detail loss in the rendered results.

\noindent \textbf{w/o Hierarchical UV Mapping.}
We disable the hierarchical pyramid and splat features onto a fixed-resolution UV map. As observed in Fig.~\ref{fig:ablation}, this results in significant blurring in areas like the nose and ears. The mismatch between the feature map resolution and the UV grid causes "holes" in the UV feature space, thereby degrading the decoding capability and expressiveness of the Gaussian primitives.

\noindent \textbf{w/o Visibility-aware Fusion.}
We remove the view-dependent weights, relying solely on UV sampling density for aggregation. As reported in Tab.~\ref{tab:ablation}, the CSIM score drops significantly, indicating a loss of identity preservation. Visually, prominent artifacts appear around the nasal bridge, where lower-quality features from oblique side views override the high-fidelity information from the frontal view. This demonstrates that visibility-aware weighting is essential for correctly prioritizing high-confidence observations.

\begin{table}[htbp]
    \centering
    \setlength{\tabcolsep}{3pt}
    \caption{\textbf{Ablation of different components on the NeRSemble dataset.}}
    \label{tab:ablation}
    \begin{tabular}{lccccc}
        \toprule
         & \multicolumn{5}{c}{\textbf{Ablation Novel Views}} \\ 
        \cmidrule(lr){2-6} 
        Method & PSNR $\uparrow$ & SSIM $\uparrow$ & LPIPS $\downarrow$ & CSIM $\uparrow$ & DS $\downarrow$ \\
        \midrule
        w/o MV Diffusion         & 22.687 & 0.774 & 0.151 & 0.655 & 0.124 \\
        w/o S.a Laplacian   & 22.988 & 0.782 & 0.150 & 0.667 & 0.105 \\
        w/o D.B Splatting   & 23.187 & 0.789 & \textbf{0.141} & 0.643 & 0.098 \\
        w/o Hierarchical UV & 22.997 & 0.785 & 0.149 & 0.655 & 0.103 \\
        w/o V.a fusion      & 23.121 & 0.793 & 0.151 & 0.597 & 0.110 \\
        Full (Ours)         & \textbf{23.221} & \textbf{0.805} & 0.144 & \textbf{0.671} & \textbf{0.095} \\
        \bottomrule
    \end{tabular}
\end{table}

\section{Conclusion and Limitations}
We presented \method{}, the first one-shot framework reconstructing generalizable, full-head, and animatable 3D avatars from a single image. By introducing a semantic-aware mesh deformation module integrating multi-view normals to optimize a haired FLAME head, and a multi-view feature splatting module aggregating full-head features into a shared canonical UV representation, our method enables effective one-shot reconstruction of complete and animation-ready 3D head avatars.
However, our framework has certain limitations. First, the underlying geometry is fundamentally constrained by the topology of the FLAME parametric model. Although we have incorporated an auxiliary teeth mesh, the model exhibits limited expressiveness for anatomical structures not explicitly modeled by FLAME, such as the tongue. Second, by explicitly binding our 3D Gaussians to the mesh surface via UV parameterization, the framework struggles to reconstruct external accessories or geometries topologically distinct from the cranium, such as eyeglasses. Addressing these topological constraints to handle complex accessories and inner-mouth details remains a promising future direction. 
With respect to ethics, to mitigate misuse risks, our framework can integrate digital watermarking and deepfake detection, promoting responsible use in telepresence and entertainment.




\bibliographystyle{ACM-Reference-Format}
\bibliography{GFGA}

\begin{figure*}[t] 
  \centering
  \includegraphics[width=0.96\textwidth, height=0.96\textheight, keepaspectratio]{
  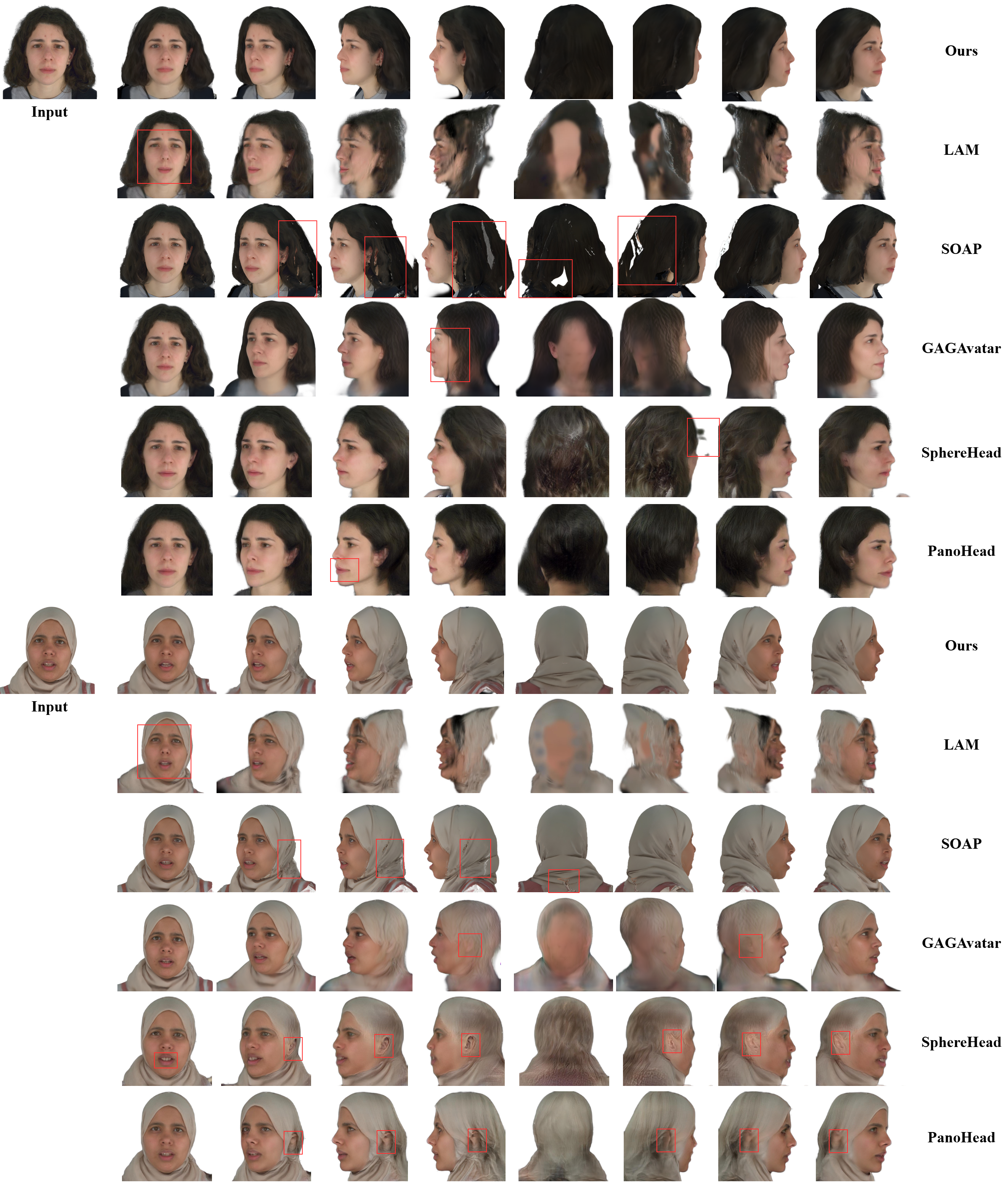}
  \caption{\textbf{More Visual results on the NeRSemble dataset.} Taking a single view as input, we perform $360^\circ$ novel view synthesis to compare our method with state-of-the-art approaches. The results show that our method achieves superior multi-view consistency and accurately reconstructs unseen regions. We use the red boxes to highlight the visual artifacts.}
  \label{fig:full-head1}
  \Description{}
\end{figure*}

\clearpage
\appendix
\section*{Supplementary Material}

In this supplementary material, we present additional results and implementation details of our method. In Section A, we elaborate on the training framework. {Section B describes the expression reenactment pipeline, presents a qualitative stress test under extreme facial expressions, and provides additional quantitative comparisons with one-shot animatable baselines.} We then describe the specific experimental settings in Section C. Section D presents additional results on novel view synthesis and $360^\circ$ full-head reconstruction.

\section{Training Details}
To achieve high-fidelity reenactment with robust geometry, our optimization is guided by a composite objective function comprising reconstruction, perceptual, and multi-view consistency terms.
First, to ensure pixel-level photometric consistency, we employ an $L_1$ reconstruction loss $\mathcal{L}_{rec}$. This is applied to both the raw splatted image ${I}_{raw}$ and the final neural-refined image ${I}_{ref}$. To further enhance high-frequency details in critical facial areas, we incorporate a bounding-box loss $\mathcal{L}_{box}$ on the cropped facial region:
\begin{equation}
\mathcal{L}_{rec} = | I_{raw} - I_{d} | + | I_{ref} - I_{d} | +\lambda_{box}\mathcal{L}_{box},
\end{equation}
To preserve high-level facial identity and structural semantics, we incorporate a perceptual loss $\mathcal{L}_{perc}$ derived from a pre-trained face recognition network~\cite{VGG-Face}.
Then, to alleviate depth ambiguity and enforce 3D geometric consistency, we introduce a pseudo-multi-view supervision $\mathcal{L}_{mv}$. We leverage the diffusion-generated multi-view data as pseudo-ground truth. Crucially, this supervision is restricted to the raw render result ${I}_{raw}$, ensuring geometric constraints while avoiding overfitting of the neural refiner to potential artifacts.
Finally, we impose regularization terms $\mathcal{L}_{local}$ on the local position offsets and scaling attributes of the UV Gaussians to prevent the UV Gaussians from drifting away from the underlying surface. The total objective function is formulated as:
\begin{equation}
\mathcal{L} = \mathcal{L}_{rec} + \lambda_{perc}\mathcal{L}_{perc} + \lambda_{mv}\mathcal{L}_{mv} + \lambda_{local}\mathcal{L}_{local},
\end{equation}
where the $\lambda$ terms are hyperparameters that balance the contributions.
To balance 3D consistency and expression flexibility, we propose a progressive side-view decay strategy. Strong multi-view constraints are essential for full-head reconstruction but hinder expression learning, while weak constraints cause frontal overfitting. We therefore start training with a high $\lambda_{mv}$ to stabilize canonical geometry, and gradually decay it to relax the constraints, enabling the avatar to learn high-frequency facial expressions while preserving underlying structure.

\noindent \textbf{Training Datasets.}
We utilize the VFHQ dataset~\cite{VFHQ} for training, which comprises high-quality video clips from various interview scenarios. To ensure data diversity and avoid consecutive redundancy, we sample 25 to 75 frames per clip depending on its length, resulting in a total of 571951 frames from 15,204 video clips. All images are resized to $512 \times 512$. Following~\cite{GAGAvatar}, we remove the background and track camera poses and FLAME~\cite{FLAME} parameters for each frame. Crucially, to bridge the gap between monocular observations and 3D supervision, we employ our pre-trained diffusion model to expand these monocular frames into pseudo-multi-view data, which provide the necessary geometric and appearance constraints for training.

\noindent \textbf{Implementation Details.}
Our framework is implemented on the PyTorch platform. We employ FLAME~\cite{FLAME} as our driving 3DMM. For feature extraction, we utilize DINOv3~\cite{DINOv3} as the backbone network, which remains frozen during the entire training process. To enable multi-view supervision, we preprocess the VFHQ dataset by generating multi-view images for each frame using a pre-trained diffusion model. These generated pseudo-multi-views are then utilized as ground truth for supervision during training. 
{The 2D neural refiner is implemented as a StyleUNet operating on the 32-channel Gaussian rasterization output. It produces a three-channel RGB image at $512 \times 512$ resolution and a lightweight StyleGAN2\cite{stylegan2} decoder. The first three channels of the rasterized feature image are treated as the raw 3D rendering, while the complete 32-channel feature image is passed to the refiner.}
We use the Adam [Kingma and Ba, 2014] optimizer with a learning rate of $1.0 \times 10^{-4}$. The model is trained for 250,000 iterations with a batch size of 6 on two NVIDIA H100 GPUs.

{
\noindent \textbf{Multi-view Generation and Mesh Deformation.}
For each source image, we use the pretrained SOAP\cite{soap} multi-view diffusion models to generate RGB images and normal maps at six yaw angles: $0^\circ$, $-90^\circ$, $180^\circ$, $90^\circ$, $-45^\circ$, and $45^\circ$. Both diffusion pipelines use a DDIM scheduler with 50 denoising steps, classifier-free guidance scale 1.5, $\eta=1.0$, and random seed 42. The RGB views are first generated at $256 \times 256$ resolution and enhanced using $4\times$ super-resolution. The resulting RGB views are then used to condition the normal-map diffusion model. 

For semantic-aware mesh fitting, we initialize the mesh from FLAME and optimize the per-vertex offsets for 800 iterations without remeshing. We use Adam with a learning rate of $1.0\times10^{-3}$. The complete fitting stage takes approximately 30 seconds per input.
}
\section{Expression Reenactment}
The goal of Expression Reenactment is to transfer the pose and expression from a driving image to the source image while preserving the source identity and maintaining high fidelity.
After decoding, the Gaussian primitives from the vertex and UV branches are combined to form a complete canonical 3D Gaussian avatar.
To animate this avatar, we first employ a face tracker~\cite{EMOCA} to extract the expression and pose parameters from the driving image. These parameters are then injected into the canonical avatar via the FLAME, explicitly driving the Gaussians to the target state.
Finally, the deformed Gaussians are rendered using the camera parameters of the driving view via splatting.

{
\noindent\textbf{Extreme-expression stress test.}
To complement the quantitative reenactment results on VFHQ\cite{VFHQ} and HDTF\cite{HDTF} reported in the main paper, we further conduct a qualitative stress test using six high-intensity driving expressions. For each source portrait in the leftmost column of Fig.~\ref{fig:extreme-expression}, we reconstruct an avatar once and transfer the expression parameters estimated from each driving image in the top row. The results show that our method captures large deformations of the mouth and eyes while preserving the identity-specific appearance of diverse source subjects. Since paired ground-truth images are unavailable for these arbitrary source--driver combinations, this experiment is presented as a qualitative stress test rather than a quantitative benchmark.

\noindent\textbf{Comparison with one-shot animatable baselines.}
We additionally compare OMEGA-Avatar with GAGAvatar~\cite{GAGAvatar} and LAM~\cite{LAM}, two strong one-shot animatable Gaussian-avatar methods. GAGAvatar and LAM primarily target frontal portrait reenactment, whereas our method additionally targets complete $360^\circ$ full-head reconstruction and animation, including side- and back-head geometry and appearance. Therefore, this comparison should be interpreted with respect to the different task scopes rather than as a strict ranking based solely on frontal reenactment quality.






\begin{table}[t]
    \centering
    \caption{
    \textbf{{Self-reenactment comparison on the VFHQ dataset.}}
    Colors denote the \first{best} and \second{second-best} results.
    }
    \label{tab:vfhq_animatable_baselines}
    \scriptsize
    \setlength{\tabcolsep}{2.2pt}
    \resizebox{\columnwidth}{!}{
    \begin{tabular}{lccccccc}
        \toprule
        \textbf{Method}
        & \textbf{PSNR$\uparrow$}
        & \textbf{SSIM$\uparrow$}
        & \textbf{LPIPS$\downarrow$}
        & \textbf{CSIM$\uparrow$}
        & \textbf{AED$\downarrow$}
        & \textbf{APD$\downarrow$}
        & \textbf{AKD$\downarrow$} \\
        \midrule

        GAGAvatar~\cite{GAGAvatar}
        & \second{21.83}
        & \second{0.818}
        & \second{0.122}
        & \second{0.816}
        & 0.111
        & 0.135
        & 3.349 \\

        LAM~\cite{LAM}
        & \first{22.65}
        & \first{0.829}
        & \first{0.109}
        & \first{0.822}
        & \first{0.102}
        & 0.134
        & \second{2.059} \\

        SOAP~\cite{soap}
        & 14.557
        & 0.599
        & 0.305
        & 0.662
        & 0.175
        & 0.0562
        & 4.997 \\

        Ours (w/o refiner)
        & 18.062
        & 0.714
        & 0.188
        & 0.664
        & 0.116
        & \second{0.0207}
        & 2.196 \\

        Ours (full)
        & 20.739
        & 0.744
        & 0.169
        & 0.660
        & \second{0.109}
        & \first{0.0188}
        & \first{2.023} \\
        \bottomrule
    \end{tabular}}
\end{table}

\begin{table}[t]
    \centering
    \caption{
    \textbf{{Self-reenactment comparison on the HDTF dataset.}}
    Colors denote the \first{best} and \second{second-best} results.
    }
    \label{tab:hdtf_animatable_baselines}
    \scriptsize
    \setlength{\tabcolsep}{2.2pt}
    \resizebox{\columnwidth}{!}{
    \begin{tabular}{lccccccc}
        \toprule
        \textbf{Method}
        & \textbf{PSNR$\uparrow$}
        & \textbf{SSIM$\uparrow$}
        & \textbf{LPIPS$\downarrow$}
        & \textbf{CSIM$\uparrow$}
        & \textbf{AED$\downarrow$}
        & \textbf{APD$\downarrow$}
        & \textbf{AKD$\downarrow$} \\
        \midrule

        GAGAvatar~\cite{GAGAvatar}
        & \second{23.13}
        & \second{0.863}
        & \second{0.103}
        & \second{0.862}
        & 0.110
        & 0.111
        & 2.985 \\

        LAM~\cite{LAM}
        & \first{23.43}
        & \first{0.873}
        & \first{0.097}
        & \first{0.865}
        & \second{0.101}
        & 0.093
        & 1.965 \\

        SOAP~\cite{soap}
        & 14.607
        & 0.711
        & 0.283
        & 0.735
        & 0.166
        & 0.0351
        & 3.579 \\

        Ours (w/o refiner)
        & 19.854
        & 0.765
        & 0.166
        & 0.757
        & 0.107
        & \first{0.0090}
        & \second{1.551} \\

        Ours (full)
        & 22.594
        & 0.820
        & 0.123
        & 0.752
        & \first{0.097}
        & \second{0.0091}
        & \first{1.515} \\
        \bottomrule
    \end{tabular}}
\end{table}






GAGAvatar\cite{GAGAvatar} and LAM\cite{LAM} demonstrate strong frontal reenactment performance, particularly in image-space fidelity and identity similarity. On VFHQ, LAM performs better in PSNR, SSIM, LPIPS, CSIM, and AED, while OMEGA-Avatar achieves lower APD and AKD. On HDTF, LAM remains stronger in image-space and identity metrics, whereas OMEGA-Avatar achieves lower AED, APD, and AKD. Meanwhile, OMEGA-Avatar additionally reconstructs and animates a complete $360^\circ$ full head, including side- and back-head geometry and appearance, which is outside the primary scope of GAGAvatar and LAM. 

{Since GAGAvatar and LAM are trained primarily with monocular frontal-view observations, their models are naturally biased toward frontal reenactment and tend to degrade under large viewpoint changes beyond the training distribution. In contrast, OMEGA-Avatar explicitly incorporates multi-view priors to support complete full-head reconstruction. These results therefore reflect a trade-off between frontal reenactment fidelity and complete $360^\circ$ full-head modeling.}
We retain SOAP as the primary animation baseline in the main paper because its full-head reconstruction setting is more closely aligned with ours.
}
\section{Experimental Details}
\noindent \textbf{Evaluation metrics}
We employ five quantitative metrics for novel view synthesis. For reconstruction quality and perceptual fidelity, we report PSNR, SSIM, LPIPS~\cite{LPIPS}, and DreamSim~\cite{DreamSim}. 
To evaluate identity preservation, we report CSIM, computing cosine similarity of face recognition features extracted using ArcFace~\cite{ArcFace}. 

For expression reenactment, we employ the Average Expression Distance (AED) and Average Pose Distance (APD) measured by a 3DMM estimator~\cite{Deep3DFace}, along with the Average Keypoint Distance (AKD) computed via a facial landmark detector~\cite{bulat2017far}. These metrics collectively evaluate expression-level and pose-level control precision during animation. 
{
Since AED and APD are computed from parameters estimated by a 3DMM regressor, their absolute values can be sensitive to the underlying 3DMM estimator, coefficient parameterization, and preprocessing. Similarly, AKD depends on the facial landmark detector and image normalization. We therefore interpret these motion-related metrics jointly rather than relying on their absolute magnitudes in isolation.
}
We evaluate our method on two publicly available multi-view datasets: NeRSemble~\cite{nersemble} and Avatar-256~\cite{avatar-256}.

\noindent \textbf{NeRSemble dataset}
The NeRSemble dataset consists of 425 identities, each captured by 16 fixed cameras. For quantitative evaluation, we randomly sampled 10 identities and selected 9 sequences for each identity. From each sequence, we further randomly sampled 50 frames. For every frame, we utilize camera 222200037 as the single input view, as it represents the dataset-defined frontal perspective, while the remaining camera views serve as the test views. The specific identities used for evaluation are listed in Tab.~\ref{tab:identities}.

\noindent \textbf{Avatar-256 dataset}
The Avatar-256 dataset consists of 256 identities, each captured by 80 cameras, with over 5,000 frames per camera. For qualitative evaluation, we randomly sampled 10 identities and selected 20 random frames from the 5,000 available frames for each identity. For each selected frame, we use camera 401168 as the single input view. This camera is selected for its frontal perspective and central position in the world coordinate system, while the remaining views are reserved for testing. The specific identities used for evaluation are listed in Tab.~\ref{tab:identities}.

For PanoHead~\cite{panohead} and SphereHead~\cite{spherehead}, we apply 3DDFA-V2~\cite{3ddfa} to align the input image to the FFHQ canonical space. The aligned image is then inverted using Pivotal Tuning Inversion~\cite{PTI} for single-view head reconstruction.

For the remaining methods, we strictly follow their official implementations to obtain the target-view camera parameters, and render the output by rotating the camera to the target view.

\section{MORE RECONSTRUCTION RESULTS}
\begin{figure*}[t] %
  \centering
  \includegraphics[width=\textwidth, height=\textheight, keepaspectratio]{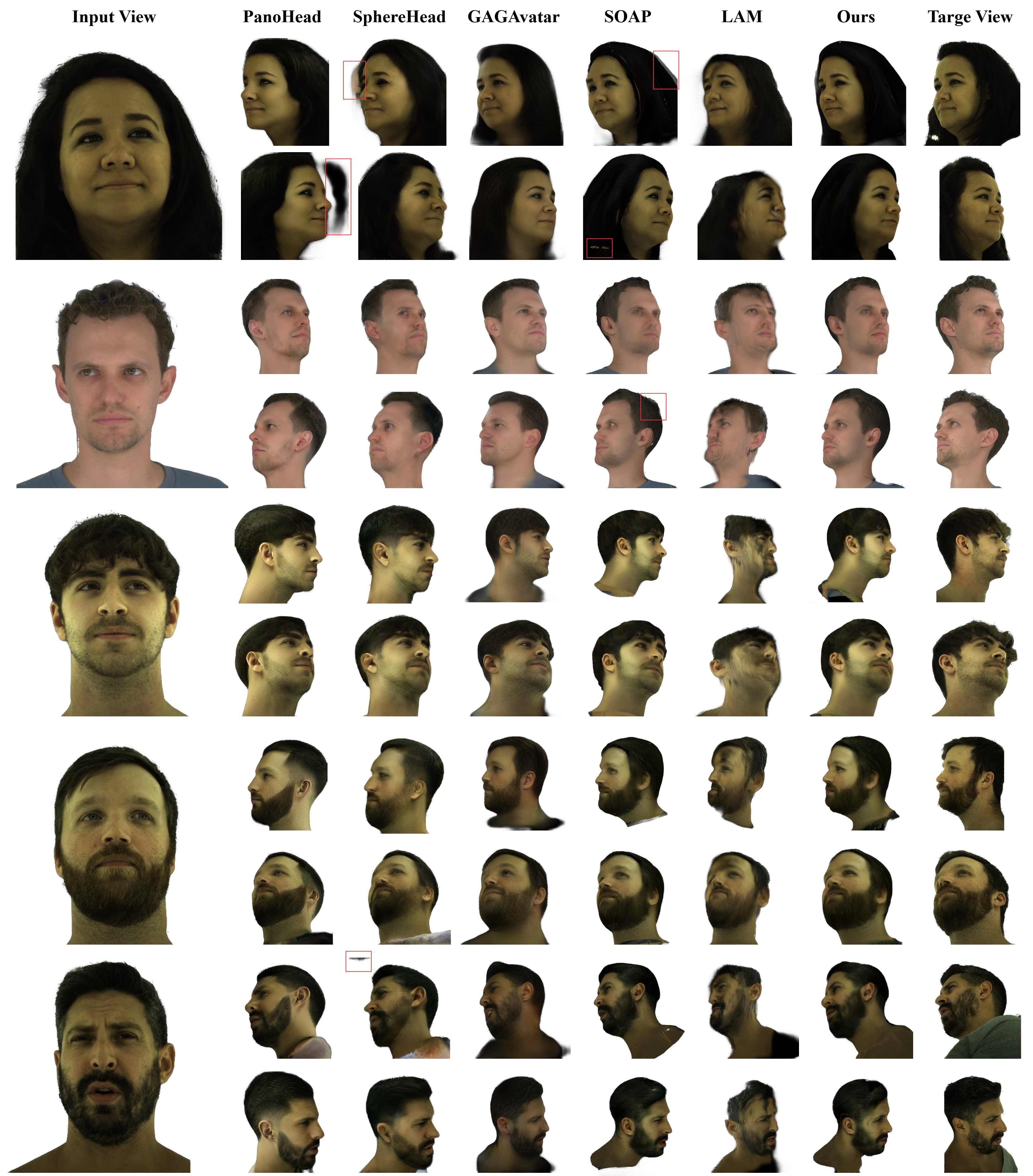}
  \caption{Additional qualitative results on the NeRSemble and Avatar-256 datasets. We use the red boxes to highlight the visual artifacts.}
  \Description{Additional qualitative results on the NeRSemble and Avatar-256 datasets}
  \label{fig:appendix}
\end{figure*}

\begin{figure*}[t]
  \centering
  \includegraphics[width=\textwidth, height=\textheight, keepaspectratio]{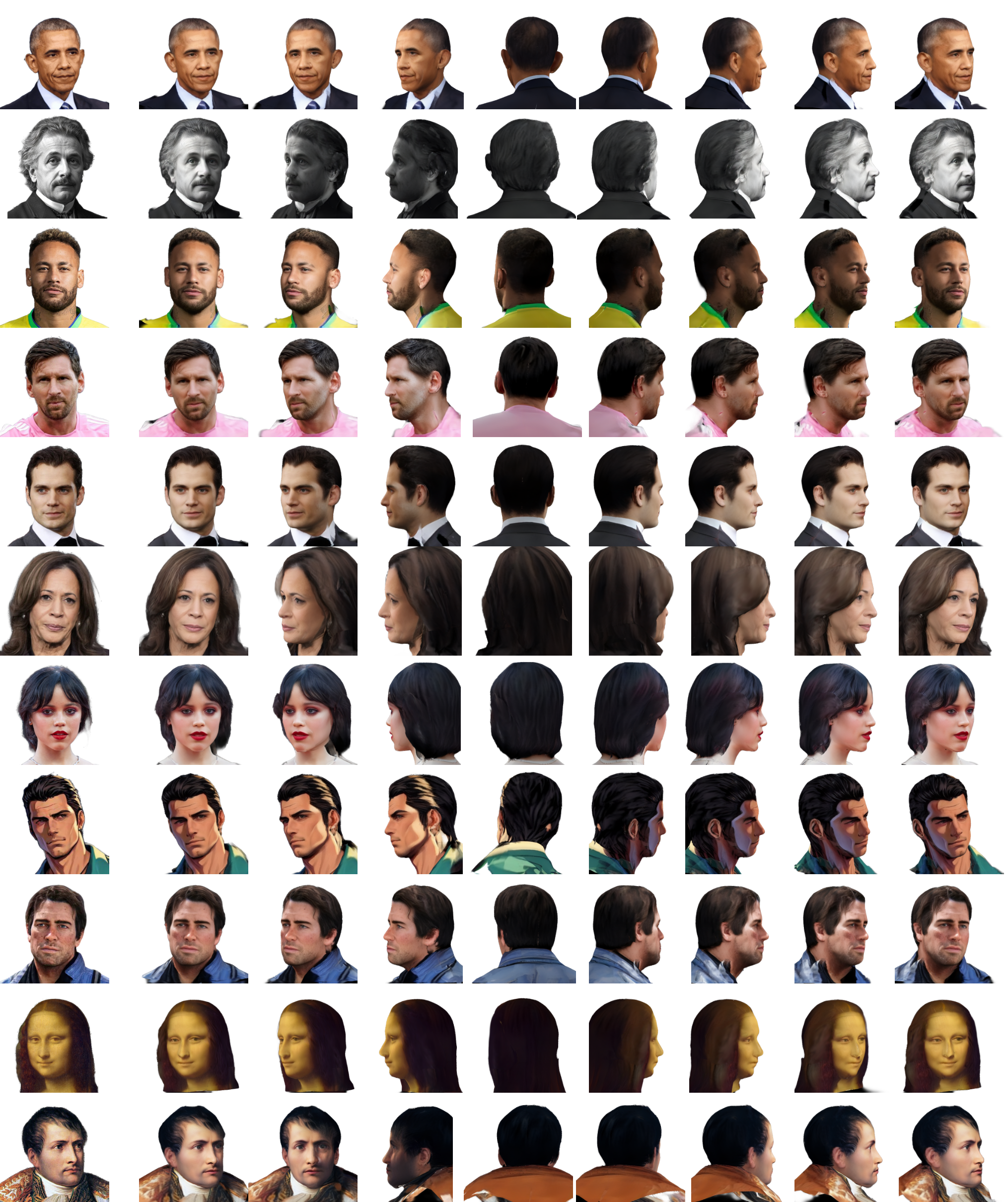}
  \caption{Full-head reconstruction results on in-the-wild images}
  \Description{Full-head reconstruction results on in-the-wild images}
  \label{fig:full-head2}
\end{figure*}

We provide additional visual results of our method. Fig.~\ref{fig:appendix} presents additional qualitative results of our method on the NeRSemble and Avatar-256 datasets for novel view synthesis, while Fig.~\ref{fig:full-head2} presents more results of our model on in-the-wild data, demonstrating $360^\circ$ full-head reconstruction.

\begin{table}[h]
    \centering
    \caption{Identities used from the NeRSemble and Avatar-256 datasets.}
    \label{tab:identities}
    \resizebox{\linewidth}{!}{ 
    \begin{tabular}{c c c}
        \toprule
        \textbf{Index} & \textbf{NeRSemble ID} & \textbf{Avatar-256 ID} \\
        \midrule
        1 & \texttt{017} & \texttt{20220614--1135--DNM410} \\
        2 & \texttt{070} & \texttt{20220808--0809--DPE040} \\
        3 & \texttt{124} & \texttt{20230308--1352--BDF920} \\
        4 & \texttt{175} & \texttt{20230328--0800--BLY735} \\
        5 & \texttt{214} & \texttt{20230405--1635--AAN112} \\
        6 & \texttt{218} & \texttt{20230726--1657--AYE877} \\
        7 & \texttt{304} & \texttt{20230728--0757--CRV122} \\
        8 & \texttt{306} & \texttt{20230810--1355--AJR151} \\
        9 & \texttt{371} & \texttt{20230901--1429--CPP930} \\
        10 & \texttt{490} & \texttt{20230908--1645--DHA971} \\
        \bottomrule
    \end{tabular}
    }
\end{table}
{
\begin{figure*}[t]
  \centering
  \includegraphics[
    width=\textwidth, height=\textheight,
    keepaspectratio
  ]{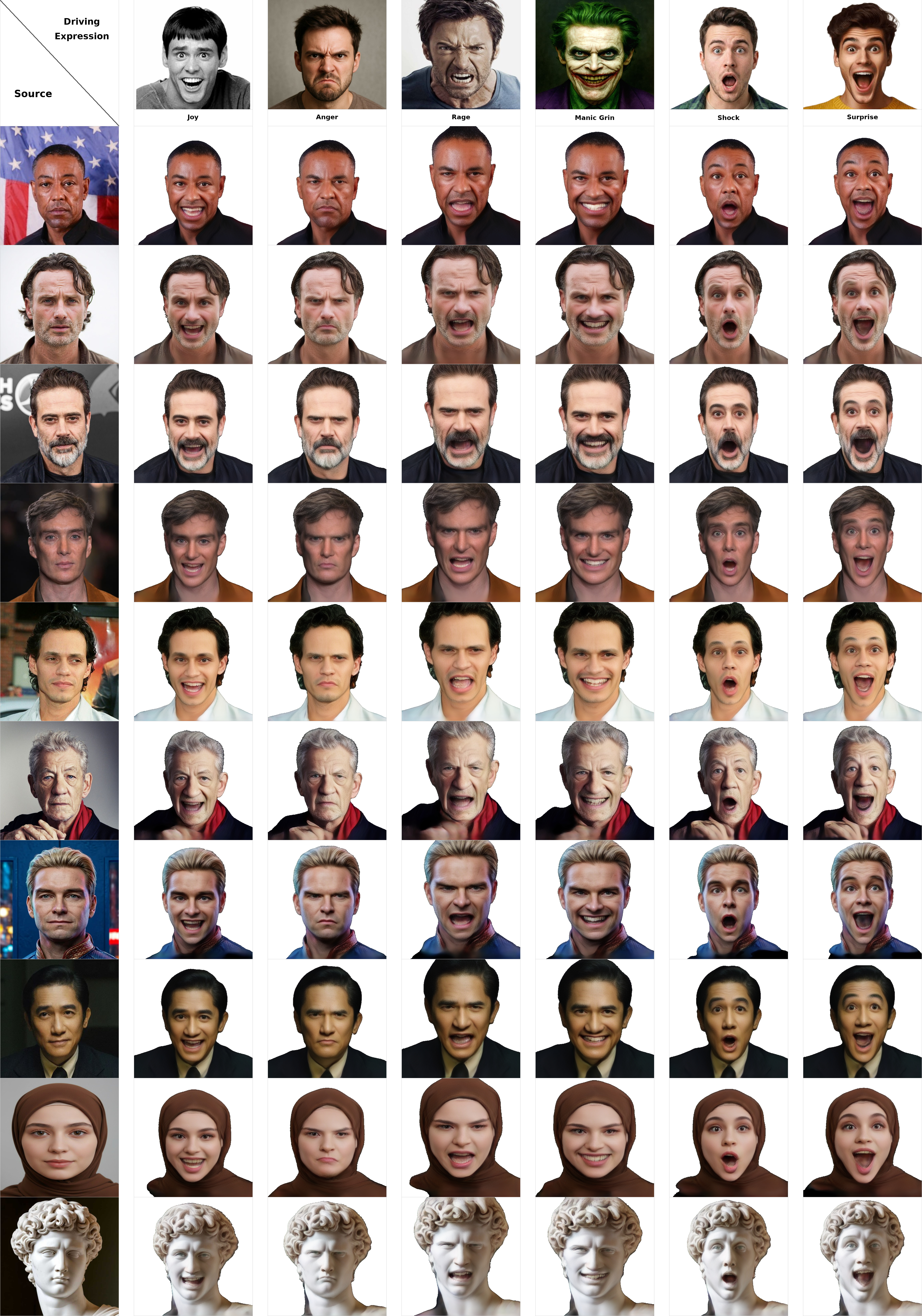}
  \caption{
    {Qualitative stress test for extreme-expression reenactment.
    The top row shows six driving images spanning joy, anger, rage,
    a manic grin, shock, and surprise, while the leftmost column shows
    the source portraits used to reconstruct the avatars. }
  }
  \Description{
    A grid with six driving expressions in the top row and ten source
    portraits in the left column. Each interior cell shows the corresponding
    source avatar reenacted with the expression of the driving image.
  }
  \label{fig:extreme-expression}
\end{figure*}

\clearpage
}

\clearpage

\end{document}